# Review of Internet Things(IoT) of Security Threats and Challenges

Fehim Köylü
*Faculty of Computer Engineering*
*Erciyes University Kayseri TURKEY*
fehimkoylu@erciyes.edu.tr

Ahmed O. Ali
*Faculty of Computer Information and Engineering Technology Zamzam University*
*Mogadishu, Somali*
geeddi114@gmail.com

Mohamud M. Hassan
*Faculty of Computer Information and Engineering Technology Zamzam University*
*Mogadishu, Somali*
mohamud@zust.edu.so

Muhiadin M. Sabriye
*Faculty of Computer Information and Engineering Technology Zamzam University*
*Mogadishu, Somali*
muhiadin@zust.edu.so
muhiadin@zust.edu.so

Abdirisak Ali Osman
*Faculty of Computer Information and Engineering Technology Zamzam University*
*Mogadishu, Somali*
abdirisaka0@gmail.com

Ali Ammar Hilal
*Faculty of Computer Engineering*
*Erciyes University Kayseri TURKEY*
ammr@gmail.com

Qazwan Abdullah
*Faculty of Electrical and*
*Electronic Engineering*
*Universiti Tun Hussein Onn*
*Malaysia*
*Johor, Malaysia*
gazwan20062015@gmail.com

*Abstract*— **In recent years, the Internet of Things (IoT) has received a lot of research attention. The IoT is considered part of the Internet of the future and is made up of billions of intelligent communication "things". The future of the Internet will consist of heterogeneously connected devices that expand the world's boundaries with physical entities and virtual components. The Internet of Things (IoT) provides new functionality for related things. This study systematically examines the definition, architecture, essential technologies, and applications of IoT. First, I will introduce various definitions of IoT. Next, it will be discussed new techniques for implementing IoT. Third, several open issues related to IoT applications will be investigated. Finally, the key challenges that need to be addressed by the research community and possible solutions to address them are investigated.**

I<small>NTRODUCTION</small>

A developing number of actual items are being associated at a remarkable rate understanding the possibility of the Internet of Things (IoT) [1]. It is the internetworking of different articles and organization networks that permit these items to convey and trade information, including sensors, brilliant meters, advanced cells, intelligent vehicles, radio-recurrence recognizable proof (RFID) labels, individual computerized associates (PDAs) and different things (inserted with gadgets, programming, and actuators) [2]. The interconnection of these gadgets empowers progressed IoT applications, e.g., item following, climate. Checking, patient's reconnaissance, and energy the executives, and extends robotization to our daily lives. One of the IoT applications is ready home, which empowers occupants to naturally open their carport when showing up home, start relaxed, plan espresso, and control lights, TV and different machines.

IoT additionally assumes an inexorably significant part in different areas, including keen city, brilliant network, e-medical services, wise transportation, mechanical robotization, and debacle reaction. It makes way for developments that work with new collaborations among "things" and humans. It gives new freedoms to applications, frameworks and administrations that improve the nature of our everyday life.

The development of IoT prompts the age of a lot of information, which has monstrous processing assets, extra room, and correspondence transfer speed. Cisco predicts that 50 billion gadgets would interface with the Internet by 2020 [3], this number would arrive at 500 billion by 2025 [4]. The information delivered by humans, machines and "things" would arrive at 500 zettabytes by 2019. However, the IP traffic of worldwide server farms would arrive at 10.4 zettabytes [5].

At that point, 45% of IoT-made information would be put away, handled, and dissected upon near or at the edge of the organization [6]. Some IoT applications may require a quick reaction. Some may include private information, which ought to be put away and designed locally. Some may deliver vast volumes of data, which could significantly weight networks [7]. Additionally, an expanding number of gadgets (e.g., keen glasses, PDAs, and vehicles) are associated with IoT for gathering and conveying fine-grained information, which may contain media data (e.g., photographs, recordings, and voices).

A lot of information brings about serious organization clog and confounded handling load on gadgets and control frameworks. With the development of IoT, mist figuring [8], [9] has been acquainted with carrying the arrangement of administrations nearer to the end-clients by pooling the accessible registering, stockpiling, and systems administration assets at the edge of the organization. It is a decentralized registering foundation that uses at least one IoT gadget or comparable client edge gadgets to cooperatively play out a considerable measure of correspondence, control, stockpiling, and the board.

Through the associations between mist hubs and gadgets, mist registering can lessen the preparing trouble on asset obliged gadgets, arrive at the idleness necessities of deferred touchy applications and beat the transfer speed imperatives for brought together administrations [10]. Mist figuring offers on-request administrations and applications general to gadgets, thick topographical appropriated and common inertness reactions, bringing about unrivaled client experience and repetition in the event of disappointment [11].

As a nontrivial augmentation of distributed computing, it is inescapable that a few issues will keep on continuing, particularly problems of security and safety [12], [13]. Mist figuring is sent by various haze specialist organizations that may not be completely trusted, and gadgets are defenseless against being undermined. Haze hubs are stood up to with additional security and protection dangers [14].

The IoT gadgets have compelled figuring, stockpiling and battery assets and are not difficult to be hacked, broken, or taken. Albeit the current arrangements in distributed computing could be relocated to address some security and protection issues in haze figuring, it has its particular security and protection challenges because of its definite highlights, like decentralized foundation, versatility support, area mindfulness and low idleness.

Then again, haze processing offers a safer foundation than distributed computing due to the nearby information stockpiling, and the non-constant information trade with cloud focuses. Mist hubs could be addressed as intermediaries for end-gadgets to perform certain tasks in the absence of the gadgets of the adequate assets [15].

Unfortunately, the security and protection issues and security assets in mist registering have not been methodically distinguished. Like this, to contemplate security and protection objectives of moisture registering is considerably essential preceding the plan and carry out of mist helped IoT applications. The exploration of the security and protection issues of haze registering for IoT is yet in its beginning phase. In this overview, it was investigated the mist helped IoT applications, security difficulties and cutting-edge arrangements.

It can be started with the advancement from cloud to mist registering, trailed by the design and highlights of mist processing. It is presented. The parts of mist hubs, including ongoing administrations, transient stockpiling, information dispersal and decentralized calculation, which add to different engaging IoT applications around there, keen home, brilliant framework, e-medical services framework, savvy transportation, and so on at that point, it will be presented the security and protection dangers and investigate the security and security challenges in mist registering.

Further, it can be surveyed the promising strategies to determine security and protection issues, break down how the current. Approaches guarantee key security objectives and ensure clients' protection in haze helped IoT applications and show our bits of knowledge on the accomplishments and leaving issues to get haze figuring. This paper is coordinated as follows. Segment 1 presents introduce the theoretical background, and Section 2 presents related work. Three security and threats. Section 4 difficulties of Internet Things (IoT). Finally, section 5 presents our conclusions and offers future research aims.

## II. RELATED WORKS

The web of things (IoT) or web of articles includes the organized interconnection of everyday items, a considerable lot of which are furnished with omnipresent knowledge [16]. A few impressive ramifications emerge from such a corpus of advancements; undoubtedly, through inserted frameworks, the IoT expands the omnipresence of the web by incorporating objects with cooperation ability [17]. In the challenge of the shrewd business [18], among the most noticeable spaces of utilization, the IoT is driving the establishment of Industry 4.0 [19], keen vehicle arrangements [20], the space of 'keen wellbeing' [21] and the advancement of savvy urban communities [22]. Organizations have been influenced on both the offer side [23]. They have subsequently presented numerous IoT-based items and administrations [24] and on the activity side [25] with significant changes to business measure frameworks [26] and the structure squares of plans of action (BMs) [27]. Accordingly, organizations' BMs have been significantly affected by the execution of IoT innovations. Nonetheless, no thorough detailed writing survey (SLR) affects BMs until this point in time.

Accordingly, the flow concentrate surveys the writing on the effect of the IoT on corporate BMs [28]; it does as such by attempted an SLR [29] utilizing the Scopus information base 'to recognize, pick and fundamentally evaluate applicable bits of exploration, and to create aggregate experiences of information from past research, It can be joined watchword examination and substance investigation to accomplish satisfactory outcomes and to respond to the accompanying examination questions (RQs) [30] 'The IoT' alludes to the organized interconnection of regular items, a considerable lot of which are outfitted with omnipresent insight [31].

While the term 'IoT' is currently generally utilized, there is no regular definition regarding what it includes. The mintage of this term is ascribed to [32], who in 1999 used the term 'web for things' in a show concerning crafted by the Auto-ID Labs at the Massachusetts Institute of Technology on organized radio-recurrence distinguishing proof framework [33].

A significant acknowledged meaning of 'IoT' is offered by the International Telecommunication Union, which characterizes it as a worldwide foundation for the Information Society, empowering progressed benefits by interconnecting (physical and virtual) things dependent on, existing and developing, interoperable data and correspondence advance. The advancement of such a corpus of innovations bears a few (and impressive) ramifications. [34]; for sure, through implanted frameworks, the IoT is expanding the pervasiveness of the web by incorporating objects with intuitive capacities [35].

This implies that the IoT has prompted a profoundly conveyed organization of gadgets that speak with individuals. Past organizations, policymakers and lead representatives are additionally turning out to be progressively mindful that the IoT is influencing abundance creation., the International Data Corporation) assessed that past 2020, the IoT will address a market worth US$7.1 trillion. Among the most unmistakable spaces of use, It will be tracked down the shrewd business, where the IoT is driving the establishment of Industry 4.0 [36], savvy transport arrangements [37], and the advancement of keen urban communities [38].

Organizations have been influenced on both the offer side [39]. They have presented numerous IoT-based items and administrations [40], and on the activity side [41], with significant alterations to business measure frameworks [42] and the structure squares of BMs [43]. During the time spent endeavor such IoT-driven alterations, a few organizations have seen significant changes in their BMs [44] talk explicitly to the shift in model structure squares and business measures, individually. A new article by [45] offers a systematization of the utilization of the IoT at four unique levels (for example, infrastructural, hierarchical, individual, and comprehensive); they layout at each level the effects on a few help areas. [46] is driven by our longing to create an independent view on the subject and keep away from the inclination for our exploration colleagues [47]. Undoubtedly, it isn't we intend to assemble an accounting technique on the side of any predefined theory.

It tends to be accepted that the SLR approach is the best independent method for revealing proof on the subject, given that 'SLR offers the chance of consolidating existing writing and making strong definitions and establishments for future examination' [48]. Organizations have been influenced on both the incentive side [39]. They have presented numerous IoT-based items and administrations [40], and on the activity side [41], with significant adjustments to business measure frameworks [42] and the structure squares of BMs [43].

During the time spent endeavor such IoT-driven alterations, the exploration group inside an SLR follows an efficient system to work with existing articles; they do as such to incorporate the information determined subsequently [49]. Such an approach needs to contain a predefined interaction by which to investigations the writing, with the goal that the cycle is reproducible. As [50] call attention to, researchers have given an assortment of interaction models that quantitatively fluctuate as far as stages and steps [51-59].

Edge registering is a comparable idea with haze figuring, which pushes the administrations, stockpiling and processing assets from focal workers to the organization edge. It can diminish the correspondence overhead between the focal and organization edges by performing information examination and information disclosure at or close to the information.

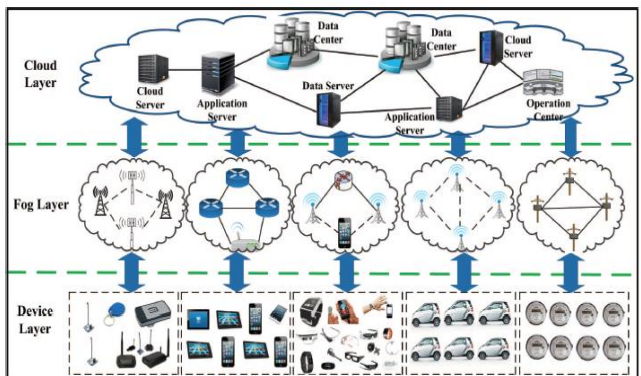

Fig. 1: Type of data collected from different applications.

Customarily, power lattice data information is communicated and traded through freely laid organization channels. Different application workers are remembered for the force lattice PC network framework, and distinctive assistance cycles and memory information bases are conveyed on every application worker. However, just the server farm and data set worker are introduced with conventional Oracle data set. The information is traded among the workers in a considerable sum and put away in the information stockpiling focus. The utilization of brought together information determined capacity makes the data set an entrance bottleneck, which seriously restricts the entrance and handling pace of information. The successive trade of a lot of communication between workers takes up a lot of organizational assets. The conventional incorporated cloud stage design requires all information to be transferred and sent after handling at the focal hub, which is frequently absurd, tedious, and wasteful.

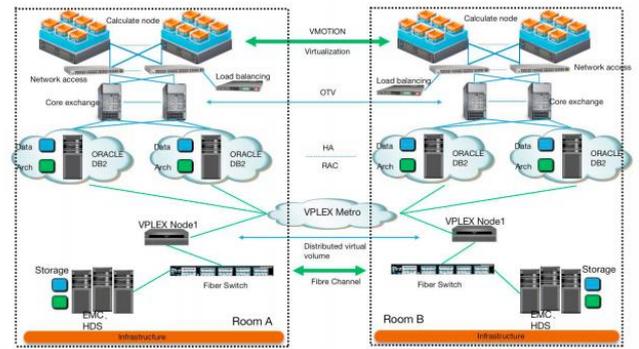

Fig. 2: Status of data center architecture.

As far as information the board, it is to construct an information and administration standard framework covering the whole chain of energy and force business on the current premise, extend the information model, and backentir company belief together administration of all information. Through the request and examination of around 90 critical business list information in six business spaces of mechanical HR the board, monetary administration, speculation project the executives, creation the board, advertising the executives, and global business, the incorporated administration of information is acknowledged, and the powerful sharing and opening of the information being framed. The record framework with the organization's essential objectives as the center backings the organization's immediate industry activity checking and normalized the executives of business information.

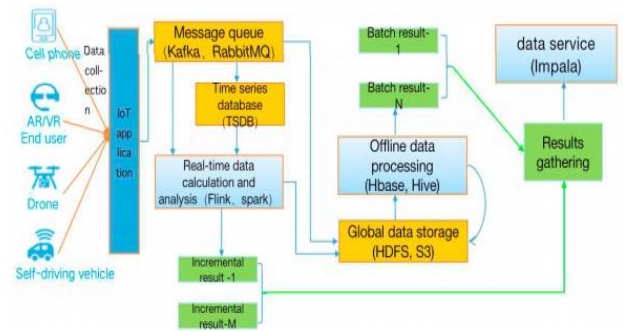

Fig. 2: Status of data centre architecture

### III. SECURITY THREATS OF INTERNET THINGS

Distributed computing is helpless against being hacked by outer assailants given the incorporated information stockpiling and registering structure. Like Google, Amazon, and Yahoo, the effectively distributed computing sellers progressively showed up enormous scope information spillage mishaps. Cloud security has become a significant factor confining the advancement of distributed computing. As a non-minor augmentation of distributed computing, haze registering is a safer engineering than distributed computing because of the accompanying reasons: First, the gathered information is fleetingly kept up and broke down on u the neighborhood mist hub nearest to information sources diminishes the reliance on the Internet associations. Local data storage, trade and examination, make it hard for programmers to access clients' information. Also, the data trade between the gadgets and the cloud no longer occurs continuously, with the goal that it is difficult for busybodies to observe the touchy data of a particular client.

Notwithstanding, mist figuring can't be considered secure since it acquires different security hazards from distributed computing. All in all, the haze hubs and mists are straightforward yet inquisitive. They are conveyed by mist and cloud sellers to offer explicit administrations sincerely to clients for their advantages. On the one hand, for financial reasons, they may not go astray from the conventions settled upon among the ones in question. Then again, they may sneak around on the substance of kept up information and the individual data about information proprietors.

Further, the representatives in mist or cloud specialist co-ops may get individual data about clients, bringing about the protection spillage for clients. Furthermore, the mist hubs or cloud workers may turn into the significant focuses of programmers that utilization any conceivable technique to arrive at their objectives deceitfully. Hence, the mist hubs or cloud workers could be straightforward yet inquisitive, even hateful. An assailant may dispatch the accompanying assaults to upset the mist processing.

Table 1: Attacks to disrupt fog computing.

| Type of threats | Descriptions |
|---|---|
| Forgery: | Malignant assailants may produce their personalities and profiles and create counterfeit data to misdirect different substances. Also, the organization assets, like transfer speed, stockpiling and energy, would be unnecessarily devoured by the faked information. |
| Tampering | altering assailants could vindictively drop, defer or adjust communicating information to disturb mist processing and debase its proficiency. |
| Spam | Spam information alludes to undesirable substances, like excess data, bogus gathered information from clients, which is created and spread by assailants. The spam would bring about pointless organization asset utilization, misdirecting social companions, and even protection spillage. |
| Collusion | At least two gatherings connive together to delude, deceive, or dupe other legitimate substances or get a little benefit. Any at least two gatherings can plan to develop their assault ability in mist noting, like a few mist hubs, IoT gadgets, IoT gadgets with the cloud, or mist hubs with IoT gadgets. |
| Man-in-the-Middle: | A malicious attacker stands or remains in two gatherings to furtively transfer or alter the trading information between these gatherings. In any case, these two gatherings accept that they are straightforwardly speaking with one another. |
| Impersonation: | A malevolent attacker imagines a genuine client to appreciate the administrations given by mist hubs or mimics a real mist hub to offer phony or phishing administrations to clients. |
| Eavesdropping: | Pernicious aggressors tune in on correspondence channels to catch communicating parcels and read the substance. This kind of organizational assault is very compelling if the information needs encryption |
| Denial-of-Service: | An attacker upsets the administrations given by mist hubs to inaccessible to its planned clients by flooding the objective haze hubs with pointless solicitations. |

III. CHALLENGE STILL FACING THE INTERNET OF THINGS

In this segment, it will be additionally discussed the security difficulties of haze registering and survey the current strategies that can be utilized to address these difficulties.

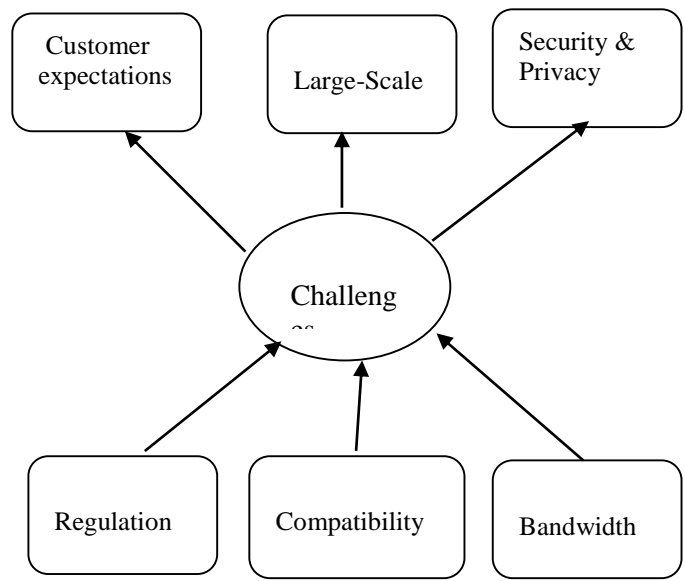

Fig. 4: Challenges and Security of Internet Things

**Regulation:** Another standard attribute of mechanical developments is that public areas guideline regularly sets aside an extended effort to find the present status of innovation. With the quick advancement that is going on consistently in IoT, the public areas are taking as much time as is needed in the making up for lost time and organizations are regularly left without essential data. They need to decide.

**Compatibility**: New influxes of innovation frequently include an enormous stable of contenders moving for a piece of the pie, and IoT is unquestionably no exemption. This can be uplifting news since rivalry makes expanded decisions for purchasers. However, it can likewise make baffling similarity issues. Home cross-section networks are one region where similarity inconvenience is approaching. Bluetooth has for quite some time been the similarity standard for IoT gadgets. Indeed, it was named after an old lord, Harald Bluetooth, known for binding together fighting clans. Be that as it may, with regards to home mechanization utilizing network organizing, a few contenders have jumped up to challenge Bluetooth's cross-section network contributions, including conventions like Zigbee and Z-Wave. It very well maybe a long time before the market settles enough to crown a solitary widespread norm for home IoT. Similarly, proceeding with similarity for IoT gadgets likewise relies on clients keeping their gadgets restored and improved, which, as it has been relatively recently talked about, can be troublesome. when IoT gadgets that need to converse with one another are running distinctive programming forms, a wide range of execution issues and security weaknesses can result. That is a significant piece of why it's imperative to such an extent that IoT buyers keep their gadgets fixed and cutting-edge.

**Bandwidth**: Network is a more significant test to the IoT than you may anticipate. As the IoT market develops IoT market size grows dramatically, a few specialists are worried that data transfer capacity concentrated IoT applications, such as video real video time, before long battle for space on the IoT's present worker customer model. That is because the worker customer model uses a concentrated worker to verify and coordinate traffic on IoT organizations. Be that as it may, as an ever-increasing number of gadgets start to interface with these organizations, they regularly battle to bear the heap. Subsequently, it's significant for IoT organizations to look at their IoT availability suppliers and pick one with a solid record of administration and development. Highlights likewise exchanging between versatile organization administrators (MNOs) are precious for making a more solid and easy-to-understand IoT item for your clients.

**Customer expectations:** It's frequently said that it's more intelligent to under-guarantee and over-convey. Numerous IoT producers have taken in this the most challenging way possible, with IoT new businesses bombing regularly and leaving confused clients afterward. hen client assumptions and item reality don't coordinate, the outcomes can be framework disappointments, stranded advancements and lost profitability. With such solid rivalry in the IoT market, clients whose assumptions aren't met will not stop for a second to go somewhere else. Organizations hoping to enter this cutthroat and the inventive area should be ready for a market that never stands by and clients who consistently need a smoother and further developed insight.

**Security & Privacy:** IoT is an energizing area with a great deal of potential to change the way it lives, work and play. Yet, the tech business, public spaces, and shoppers are the same, and shoppers should agree about safety and execution issues to guarantee that the IoT stays protected and beneficial to utilize.

## IV. CONCLUSION

This article presents an outline of IoT and another decentralized design that alters distributed computing by broadening stockpiling, registering, and organizing assets to the organization edge to support incredibly huge scope IoT applications. It is also faced with customary security dangers, which raise different new security and protection challenges towards clients. In this article, it has been given an extensive review of getting haze processing for IoT applications. It has been first inspected the engineering and the highlights of haze registering. It has also been discussed the jobs of mist hubs in IoT applications, including continuous administrations, transient stockpiling, information scattering, and decentralized calculation, and inspected a few promising IoT applications as per various parts of mist hubs. security and protection dangers in mist processing have been introduced, including a progression of safety assaults and protection openness hazards. Move over, it has been exhibited the security and protection challenges, audited the condition of-workmanship answers for secure haze registering in IoT applications, and showed our experiences on the leaving issues to further examine the security and protection issues. At last, it has been recognized a few open examination issues able that should contain the most reason for security and protection issues in mist processing.

## Acknowledgment

The authors would like to thank Zamzam University of Science & Technology (Zust) for sponsoring this work.

support for IoT based product-service systems (PSS). Business Process Management Journa